\documentclass[12pt,a4paper]{article}
\usepackage{graphicx}
\usepackage[cp1251]{inputenc}
\usepackage[english]{babel}
\usepackage{amssymb, amsfonts, amsmath}
\usepackage{amscd}
\begin{document}

\date{}

\title{Exact solutions of Maxwell equations in homogeneous spaces with the group of motions $G_3(VIII)$}

\author{V. V. Obukhov}

\maketitle

\quad Institute of Scietific Research and Development,
Tomsk State Pedagogical University (TSPU), 60 Kievskaya St., Tomsk, 634041, Russia; \\ \quad

Laboratory for Theoretical Cosmology, International Center of Gravity and Cosmos, Tomsk State University of Control Systems and Radio Electronics (TUSUR), 36, Lenin Avenue, Tomsk, 634050, Russia

\quad


\section{Introduction}
If the symmetry of space-time and physical fields is given by Killing fields whose number is not less than three, it is possible to reduce the field equations and the equations of motion of the tested charged particles to the systems of ordinary differential equations.

The spaces admitting complete sets of mutually commutative Killing tensor fields of rank not greater than two are of special interest in the theory of gravitation. Such spaces are called Steckel spaces. The theory of Steckel spaces was developed in \cite{1}, \cite{2}, \cite{3}, \cite{4}, \cite{5}, \cite{6}, \cite{7}. (see also \cite{8},\cite{9} \cite{10}, \cite{11}, and the bibliography is given there). The equations of motion of test particles in Stackel spaces can be integrated using the commutative integration method - CIM (or the method of complete separation of variables). Exact solutions of the gravitational equations are still actively used in the study of various aspects of gravitational theory and cosmology (see, for example, \cite{12}, \cite{13}, \cite{14}, \cite{15}, \cite{15a}, \cite{15b},\cite{15c}, \cite{16}, \cite{17}, \cite{17a}, \cite{18}, \cite{19}.

Another method for exact integration of the equations of motion for a test particle (the method of non-commutative integration (NCIM)) was proposed in \cite{23}. The method is applied to spaces admitting non-commutative groups of motion \quad $G_r (r) ,r\geq 3$ \quad (see A. Petrov \cite{23a}).
It allows for reducing the equations of motion to systems of ordinary differential equations. By analogy with Stackel spaces, we call them poststack spaces - PSS. PSS are also actively studied in gravitational theory and cosmology (see, e.g., \cite{24}, \cite{25}, \cite{26}, \cite{20o}, \cite{21o}, \cite{22o}, \cite{22A}, \cite{22B}).
The classification of electromagnetic fields in which the Klein-Gordon-Fock equations and Hamilton-Jacobi equations admit non-commutative algebras of symmetry operators for a charged sample particle is carried out in \cite{28}, \cite{29}, \cite{30}, \cite{31},

The commutative and non-commutative integration methods have a similar classification problem, namely enumerating all non-equivalent metrics and electromagnetic potentials satisfying the requirements of the given symmetry. For Stackel spaces, the problem of classifying admissible external electromagnetic fields and electrovacuum solutions of the Einstein-Maxwell equations was solved in \cite{29a}.

In the previous works (\cite{34}, \cite{35}, \cite{36}), the non-null PSS of all types were considered according to the Bianchi classification except the type $VIII$. In the present work, all non-equivalent exact solutions of Maxwell's vacuum equations for non-null PSS of type $VIII$ are obtained. Thus this classification is completed for all non-null PSS.

\section{Admissible  electromagnetic fields in homogeneous spaces}

According to the definition (see \cite{32}) the space-time \quad$V_4$\quad is homogeneous if its metric can be represented in a semi-geodesic coordinate system as follows:
\begin{equation}\label{1}
ds^2 = -{du^0}^2 + \eta_{ab}e_{\alpha}^a e_{\beta}^b du^\alpha du^\beta,  \quad g_{ij}=-\delta_i^0\delta_j^0 + \delta_i^\alpha\delta_j^\beta e^a_\alpha e^b_\beta \eta_{ab}(u^0), \quad det|\eta_{ab}|=\eta^2 >0, \quad \dot{e}^a_{\alpha}=0,
\end{equation}
and the condition
\begin{equation}\label{2a}
[Y_a,Y_b] = C_{ab}^c Y_c, \quad Y_a= e_a ^\alpha \hat{\partial}_\alpha
\end{equation}
is satisfied. Here \quad $e_a ^\alpha$ \quad are the triad of the dual vectors:
\begin{equation}\label{1f}
  e^b_\alpha e_a ^\alpha =\delta_a^b.
\end{equation}
$C^a_{bc}$ \quad are structural constants of the group \quad $G_3(N)$,\quad which acts on \quad $V_4$.
The vectors of the frame \quad $e^a_\alpha$ \quad define a non-holonomic coordinate system in the hypersurface of transitivity  \quad $ V_3$ \quad of the group \quad $G_3(N)$.\quad Here and elsewhere, dots denote the derivatives on the variable \quad $u^0$.\quad
The coordinate indices of the semi-geodesic coordinate system are denoted by letters:\quad $i, j, k = 0, 1 \dots 3$.\quad The variables of the local coordinate system on   \quad $ V_3$ \quad are provided with indices:\quad $\alpha, \beta, \gamma=1, \dots 3.$\quad Indices of non-holonomic frame are provided with indices:\quad $a, b, c = 1, \dots 3 $.\quad The rule is used, according to which the repeating upper and lower indices are summarized within the index range.


It has been proved in the paper \cite{29}, that  for a charged test particle moving in the external electromagnetic field with potential $A_i$, the Hamilton-Jacobi equation:
\begin{equation}\label{4}
g^{ij}(p_i+A_i)(p_j+A_j)=m^2 \quad (p_i+A_i = P_i),
\end{equation}
and the Klein-Gordon-Fock equation:
\begin{equation}\label{2f}
\hat{H}\varphi = (g^{ij}(-i\hat{p}_l+A_l)(-i\hat{p}_j+A_j)=m^2\varphi \quad (-i\hat{p}_j+A_j =\hat{P}_j)
\end{equation}
admit the integrals of motion
\begin{equation}\label{3f}
X_\alpha=\xi_\alpha^i p_i \quad (or \quad \hat{X}_\alpha=\xi_\alpha^i \hat{p}_i),
\end{equation}
if and only if the conditions:
\begin{equation}\label{8}
\xi^\alpha_a(\xi^\beta_b A_\beta)_{,\alpha}= C^c_{ab}\xi^\beta_b A_\beta \end{equation}
are satisfied.
Here \quad $ p_i=\partial_i\varphi $; \quad $\hat{p}_k=-\imath\hat{\nabla}_k;\quad  $  ($ \hat{\nabla}_k$ is the covariant derivative operator corresponding to the partial derivative operator -\quad   $\hat{\partial}_i,$   \quad  $\varphi $\quad is a scalar function of the particle with mass $\quad m $); \quad  $\xi_\alpha^i$ is Killing vectors,\quad   $C^c_{ab}$ are structural constants:
$$[\hat{X}_a,\hat{X}_b] = C_{ab}^c \hat{X}_c.$$

If $A_i$ satisfies condition \eqref{8}, the electromagnetic field is called admissible.
All admissible electromagnetic fields for groups of motion \quad $G_r(N)\quad ( r\geq 3)$ \quad acting transitively on hypersurfaces of the space-time have been found in \cite{29}, \cite{30}, \cite{31}.

Let us show that solutions of the system of equations \eqref{8} for HPSS of type $VIII$ can be represented in the form:
\begin{equation}\label{8a}
A_{\alpha}=\alpha_a(u^0) e_{\alpha}^a  \Rightarrow \mathbf{A}_a = e^{\alpha}_a A_\alpha = \alpha_a(u^0).
\end{equation}
To prove this, let us find the frame vector using the metric tensor of Bianchi's $VIII$-type space (see \cite{23a}).
\begin{equation}\label{4f}
{ds}^2 = {du^1}^2a_{11}+2{du^1du^2}(a_{11}{u^{1}}^2-
2a_{13}u^{1} + a_{12})\exp(-u^3)+2{du^1du^3}(a_{13}-a_{11}u^{1})+
\end{equation}
$$
+{du^2}^2(a_{11}{u^{1}}^4-4a_{13}{u^1}^3 + 2(a_{12}+2a_{33}){u^1}^2 -4a_{23}{u^1}+a_{22})\exp(-2u^3)
$$
$$
2{du^2du^3}(-a_{11}{u^{1}}^3+3a_{13}{u^1}^2 - 2(a_{12}+2a_{33}){u^1} +a_{23})\exp(-u^2)+{du^3}^2(a_{11}{u^{1}}^2-2a_{13}{u^1} + a_{33})+
\varepsilon{du^0}^2.$$
$a_{ab}$  are arbitrary functions on $u^0, \quad \varepsilon^2 = 1$.

To obtain the functions \quad $e^\alpha_a$ \quad it is sufficient to consider the components \quad $ g_{11},\quad  g_{12},\quad g_{13} $ \quad from the system \eqref{1}. The solution can be represented in the form:
\begin{equation}\label{1c}
 e^a_\alpha=\begin{pmatrix} 1 & 0  & 0 \\{u^1}^2 \exp(-u^1) & \exp(-u^3) & -2{u^1} \exp(-u^3) \\-u^1 & 0 & 1\end{pmatrix},
 e_a^\alpha=\begin{pmatrix} 1 & 0  & 0 \\{u^1}^2 & \exp(u^3) & 2{u^1} \\u^1 & 0 & 1\end{pmatrix}.
\end{equation}

The lower index numbers the lines. The solution of the system of equations \eqref{8} has been found in \cite{29}. It has the form:
$$
A_1=\alpha_0(u^0), \quad A_2 = (\alpha_0 {u^1}^2 +2\beta_0(u^0)u^1 +\gamma_0(u^0)), \quad A_3=-(\alpha_0 {u^1} +\beta_0).
$$

By denoting:\quad $\alpha_0 = \alpha_1, \quad \gamma_0 = \alpha_2, \quad \beta_0 = -\alpha_3,$ \quad we get \eqref{8a}.

\section {Maxwell's equations}

All exact solutions of empty Maxwell's equations for solvable groups have been found in the papers \cite{34}, \cite{35}. The present paper solves the problem for the group  $G_3(VIII)$.

Consider empty Maxwell's equations for an admissible  electromagnetic field in homogeneous space with a group of motions $G_r$:
\begin{equation}\label{13}
\frac{1}{\sqrt{-g}}(\sqrt{-g}F^{ij})_{,j} = 0,
\end{equation}

The metric tensor and the electromagnetic potential are defined by relations \eqref{1}, \eqref{8a}.
When $i=0$, from the set of equations \eqref{13} it follows:

\begin{equation}\label{14}
\quad\frac{1}{\sqrt{-g}}(\sqrt{-g}F_{0.}^{.\alpha})_\alpha = \frac{1}{e}( e^{\alpha}_{a}e\eta^{ab}\dot{\alpha}_{b})_{,\alpha} = \rho_a\frac{(\eta^{ab}\eta\dot{\alpha}_b)}{\eta} =0 \quad (\rho_a  = e^\alpha_{a,\alpha} + e^\alpha_a e_{,\alpha}/e).
\end{equation}
Here it is denoted:
$$
g=-\det||g_{\alpha\beta}||=-(\eta e)^2, \quad where \quad \eta^2 = \det||\eta_{\alpha\beta}||, \quad e=\det||e_{\alpha}^a||. \quad
$$
Let \quad $i=\alpha$. \quad Then, from equation \eqref{13}, it follows:
\begin{equation}\label{16}
\frac{1}{\eta}(\eta F_{0.}^{.\alpha})_{,0} = \frac{1}{e}(eF^{\beta \alpha})_{,\beta} \Rightarrow \frac{1}{\eta}(\eta \eta^{ab}e^\alpha_a \dot{\alpha}_{b})_{,0}=\frac{1}{e}(e^\beta_b \eta^{ab}e_{\tilde{a}}^{\alpha} e_{\tilde{b}}^{\gamma} \eta^{\tilde{a}\tilde{b}}F_{\beta\gamma}e e_a^\nu)_{,\nu}\Rightarrow
\end{equation}
\begin{equation}\label{16a}
e(\dot{\alpha}_b\eta\eta^{ab})_{,0} = \eta e_\alpha^a(e e^\beta_b e_{\tilde{a}_1}^{\alpha} e_{\tilde{b}}^{\gamma}F_{\beta\gamma})_{|{a_1}}\eta^{{a_1}b}\eta^{\tilde{a}\tilde{b}}.
\end{equation}
Let us find components of $F_{\alpha\beta}$ using  the relations \eqref{8a}.
\begin{equation}\label{17}
F_{\alpha\beta} = (e^a_{\beta,\alpha}-e^a_{\beta,\alpha})\alpha_a = e^c_\beta e^\gamma_c e^d_\alpha e^\nu_d(e^a_{\gamma,\nu}-e^a_{\nu,\gamma})\alpha_a=e^b_\beta e^a_\alpha e^c_\gamma  (e^\gamma_{a|b}-e^\gamma_{b|a})\alpha_c = e^b_\beta e^a_\alpha C^c_{ba}\alpha_c.
\end{equation}
Then
\begin{equation}\label{18}
(eF^{\alpha\beta})_{,\beta} = \eta^{ab}\eta^{\tilde{a}\tilde{b}}C^d_{\tilde{b}b}\alpha_d ((ee^\alpha_a)_{|\tilde{a}} + ee^\alpha_ae^\gamma_{\tilde{a},\gamma}).
\end{equation}
We present the structural constants of a  group $G_3$ in the form:
\begin{equation}\label{19}
C^c_{ab} = C^c_{12}\varepsilon^{12}_{\tilde{a}\tilde{b}} + C^c_{13}\varepsilon^{13}_{\tilde{a}\tilde{b}} + C^c_{23}\varepsilon^{23}_{\tilde{a}\tilde{b}},
\end{equation}
where
$$
\varepsilon^{AB}_{ab} = \delta^A_a\delta^B_b - \delta^A_b\delta^B_a.
$$
Let us denote:
$$
\sigma_1 = C^a_{23}\alpha_a, \quad \sigma_2 = C^a_{31}\alpha_a, \quad \sigma_3 = C^a_{12}\alpha_a;
$$
$$
\left\{\begin{array}{rl}
\gamma_1=\sigma_1\eta_{11}+\sigma_2\eta_{12}+\sigma_3\eta_{13}, \\
\gamma_2=\sigma_1\eta_{12}+\sigma_2\eta_{22}+\sigma_3\eta_{23}, \\
\gamma_3=\sigma_1\eta_{13}+\sigma_2\eta_{23}+\sigma_3\eta_{33}. \\
\end{array}\right.
$$
Equations \eqref{18} will take the form:
\begin{equation}\label{19}
\eta(\eta\eta^{ab}\dot{\alpha}_b)_{,0} = \delta^a_1(\gamma_1(C^1_{32}) - \gamma_2(C^1_{31} +\rho_3) + \gamma_3(C^1_{21} +\rho_2)) + \delta^a_2(\gamma_1(C^2_{32} +\rho_3) +
\end{equation}
$$
\gamma_2C^2_{13} - \gamma_3(C^2_{12} +\rho_1)) +\delta^a_3(-\gamma_1(C^3_{23} +\rho_2) + \gamma_2(C^3_{13} +\rho_1) + \gamma_3C^3_{21}),
$$
\begin{equation}\label{19g}
\rho_a \eta^{ab}\dot{\alpha}_b =  0.
\end{equation}
To decrease the order of the equations \eqref {19}, we introduce new independent functions:
\begin{equation}\label{20}
 b^a = \delta^c_a b_c = \eta\eta^{ab}\dot{\alpha}_b \quad \Rightarrow \quad\eta \dot{\alpha}_a = \eta_{ab}b^b.
\end{equation}
Let us introduce the function:
\begin{equation}\label{19f}
n_{ab}=n_{ab}(u^0)=\frac{\eta_{ab}}{\eta} \quad \Rightarrow \quad\det{n_{ab}} = n=\frac{1}{\eta}.
\end{equation}
Then Maxwell's equations \eqref{19} and \eqref{19f} take the form of a system of linear algebraic equations on the unknown functions $n_{ab}$:
\begin{equation}\label{20f}
 \dot{b}^{a} = \delta^a_1(\tilde{\gamma}_1(C^1_{32}) - \tilde{\gamma}_2(C^1_{31} +\rho_3) + \tilde{\gamma}_3(C^1_{21} +\rho_2)) + \delta^a_2(\tilde{\gamma}_1(C^2_{32} +\rho_3) +
\end{equation}
$$\tilde{\gamma}_2 C^2_{13} - \tilde{\gamma}_3(C^2_{12} +\rho_1)) +\delta^a_3(-\tilde{\gamma}_1(C^3_{23} +\rho_2) + \tilde{\gamma}_2(C^3_{13} +\rho_1) + \tilde{\gamma}_3C^3_{21}) \quad (\tilde{\gamma}_a = n\gamma_a),$$
\begin{equation}\label{20}
 \dot{\alpha}_a = n_{ab}b^b.
\end{equation}
The equation \eqref{19g}:
\begin{equation}\label{21}
\rho_a b^a = 0
\end{equation}
is a restriction on the function $b^a$ (if $\rho_a \ne 0$).
Let us obtain  the  Maxwell's equations for the group $G_3(VIII)$. Non-zero  structural constants, in this case, have the form:
\begin{equation}\label{21f}
C^3_{12} =2, \quad C^1_{13}=1, \quad C^2_{32} = 1 \Rightarrow
\end{equation}
From here, it follows:
$$
\sigma_1 =-\alpha_2, \quad \sigma_2 = -\alpha_1, \quad \sigma_3 = 2\alpha_3. \quad
$$
Using these relations, we obtain Maxwell's equations \eqref{19} -   in the form:
\begin{equation}\label{2c}
\hat{B}\hat{n} = \hat{\omega},
\end{equation}
where
\begin{equation}\label{3c}
\hat{B}=\begin{pmatrix} a_1 & a_2 & a_3 & 0 & 0 & 0
\\b_1 &b_2 &b_3 & 0 & 0 & 0 \\ 0 & a_1 & 0 &a_2 &a_3 & 0 \\ 0 & b_1 & 0 &b_2 &b_3 & 0 \\ 0 & 0 & a_1 & 0 & a_2 &a_3
\\ 0 & 0 & b_1 & 0 & b_2 & b_3 \end{pmatrix},
\end{equation}
$$
\hat{n}^T = ( n_{11}, n_{12}, n_{13}, n_{22}, n_{23}, n_{33}) ; \quad
\hat{\omega}^T = (-\dot{b}_{2}, \dot{a}_{2}, -\dot{b}_{1}, \dot{a}_{1}, \frac{\dot{b}_{3}}{2}, -\frac{\dot{a}_{3}}{2}).
$$
Here and after next notations are used:
\begin{equation}\label{3cc}
\alpha_1=a_2, \quad \alpha_2=a_1, \quad \alpha_3=-\frac{a_3}{2}.
\end{equation}
Let us find the algebraic complement of the matrix $\hat{B}:$
\begin{equation}\label{4c}
\hat{V}=\begin{pmatrix} b_1 v^2_1 & -a_1v^2_1 &b_2 v^2_1 &-a_2 v^2_1 & b_3 v^2_1 & -a_3V^2_1
\\b_1v_1v_2 & -a_1v_1v_2 &b_2 v_1v_2 &-a_2v_1v_2 & b_3 v_1v_2 & -a_3v_1v_2\\b_1 v_1v_3 & -a_1v_1v_3 &b_2 v_1Vv_3 &-a_2v_1v_3 & b_3 v_1v_3 & -a_3v_1v_3 \\ b_1 v^2_2 & -a_1v^2_2 &b_2 v^2_2 &-a_2v^2_2 & b_3 v^2_2 & -a_3v^2_2 \\b_1 v_2v_3 & -a_1v_2v_3 &b_2 v_2v_3 &-a_2v_2v_3 & b_3 v_2v_3 & -a_3v_2v_3
\\b_1 v^2_3 & -a_1v^2_3 &b_2 v^2_3 &-a_2v^2_3 & b_3 v^2_3 & -a_3V^2_3\end{pmatrix}
\end{equation}
$$
v_1=a_2b_3 - a_3b_2, \quad v_2=a_3b_2 - a_2b_3, \quad v_3=a_1b_2 - a_2b_1, \quad
$$
As $\hat{B}$ is a singular matrix, $\hat{V}$ is the annulling matrix for $\hat{B}$:
\begin{equation}\label{24f}
\hat{V}\hat{B}=0.
\end{equation}
So when \quad ${v_1}^2 +{v_2}^2 + {v_3}^2 \ne 0$,\quad one of the equations from the system \eqref{2c} can be replaced by the equation:
\begin{equation}\label{25f}
{a_3}^2 + {b_3}^2 = 4({a}_1{a}_2 + {b}_1{b}_2 + c) \quad (c =const).
\end{equation}
Depending on the rank of the matrix $\hat{B}$, one or more functions \quad $n_{ab}(u^0)$ \quad  are independent.
It is possible to express the remaining functions \quad $n_{ab}$ \quad  through the functions \quad $a_a, b_a$.
To find non-equivalent solutions of the system \eqref{2c}, one should consider the following variants:

$1.\quad a_1\ne 0; \quad 2.\quad a_1= 0, \quad a_2\ne 0; \quad 3.\quad a_1 = a_2 = 0, \quad a_3\ne 0.$ \quad  Taking this observation into account, let us consider all non-equivalent options.

\section{Solutions of Maxwell equations}

Since the functions \quad $a_a$ \quad satisfy the condition:
$$
a_1^2 + a_2^2 + a_3^2 \ne 0,
$$
$rank$ of matrix \eqref{4c} cannot be less than 3. If
$$
v_1^2 + v_2^2 + v_3^2 \ne 0 \Rightarrow rank||\hat{B}||=5.
$$
In order to obtain a complete solution to the classification problem, it is necessary:

I\quad to consider all non-equivalent variants with non-zero minors of \quad $rank = 5$ \quad  of the matrix \quad  $\hat{B}$;

II\quad to consider all non-equivalent variants under the condition: \quad
$v_a=0 \quad  (rank \leq 3).$

The components of the matrix \quad  $\hat{\eta}$ \quad  as well as the functions \quad  $\alpha_a$ \quad  are given by formulae  \eqref{19f}, \eqref{3cc}
In view of these circumstances, let us list all exact solutions of empty Maxwell equations for PSS of type VIII.

\quad

I. \quad $rank||\hat{B}||=5$.

1.  \quad $a_1v_1 \ne 0 \Rightarrow$ the minor $\hat{B}_{12}$ and its inverse matrix $\hat{P}=\hat{B}_{12}^{-1}$ have the form:
\begin{equation}\label{6c}
\hat{B}_{12}=\begin{pmatrix} a_2 & a_3 & 0 & 0 & 0
 \\ a_1 & 0 &a_2 &a_3 & 0 \\ b_1 & 0 &b_2 &b_3 & 0 \\ 0 & a_1 & 0 & a_2 &a_3
\\0 & b_1 & 0 &b_2 & b_3 \end{pmatrix},
\end{equation}
\begin{equation}\label{7.c}
\hat{P}=\begin{pmatrix}-\frac{v_2}{a_1v_1} & -\frac{a_3b_2 }{\alpha_1v_1} & \frac{a_2 a_3 }{a_1v_1} & -\frac{a_3b_3}{a_1v_1} & \frac{a^2_3}{a_1v_1}
 \\
-\frac{V_3}{a_1v_1} & \frac{a_2b_2}{a_1v_1} & -\frac{a_2^2 }{a_1v_1} &\frac{a_2 b_3}{a_1v_1} & -\frac{a_2a_3}{a_1v_1}
\\
-\frac{V_2^2}{a_1v_1^2} &\frac{(a_3 b_1v_1 -a_2 b_3v_3)}{a_1v_1^2} & \frac{a_3(a_2 v_2-a_1 v_1)}{a_1v_1^2} &-\frac{a_3b_3 v_2}{a_1v_1^2} & \frac{a_2^2v_2}{a_1v_1^2}
\\
 -\frac{v_2v_3 }{a_1v_1^2} & \frac{a_2b_2v_2}{a_1v_1^2} &-\frac{a^2_2 V_2}{a_1v_1^2} &-\frac{a_3b_3v_3}{a_1v_1^2} & \frac{a_3^2 v_3}{a_1v_1^2}
\\
-\frac{v_3^2}{a_1v_1^2} & \frac{a_2b_2v_3}{a_1v_1^2}&-\frac{a^2_2 v_3^2 }{a_1v_1^2}&\frac{(a_3 b_2v_3-a_2 b_1v_1)}{a_1v_1^2} & \frac{a_2 (a_1v_1 - a_3 v_3)}{a_1v_1^2}\end{pmatrix}
\end{equation}
Then the solution of equation \eqref{2c} is as follows:
\begin{equation}\label{8c}
\hat{n}_1 = \hat{P}_1\hat{\omega}_1,
\end{equation}
were
$$
\hat{n}_1^{T} = (n_{12}, n_{13}, n_{22}, n_{23}, n_{33}) ; \quad
$$
$$
\hat{\omega}_1^{T} = (-(\dot{b}_{2}+a_1n_{11}), -\dot{b}_{1}, \dot{a}_{1}, \frac{\dot{b}_{3}}{2}, -\frac{\dot{a}_{3}}{2}).
$$
Function \quad $n_{11},$ \quad  $a_a, \quad b_a$ \quad are arbitrary functions of \quad $u^0$, \quad that obey the condition \eqref{25f}.

\quad

2. \quad $a_2v_1 \ne 0 $. Obviously, we obtain a non-equivalent solution to the previous one only if \quad  $a_1=0$.\quad  In order to implement the classification, a similar choice should be made for all other variants. The matrix \quad  $\hat{B}_{14}$ \quad  and its inverse matrix \quad  $\hat{P}_2 = \hat{B}^{-1}_{14}$ \quad  have the form:
\begin{equation}\label{9c}
\hat{B}_{14}=\begin{pmatrix} a_2 & \alpha_3 & 0 & 0 & 0
 \\ b_2 & b_3 & 0 & 0 & 0 \\ 0 & 0 & a_2 & a_3 & 0 \\ 0 & 0 & 0 &a_2 & a_3
\\0 & b_1 & 0 &b_2 & b_3 \end{pmatrix},  \quad
\hat{P}_2=\begin{pmatrix}
 \frac{b_3}{ v_1}  & -\frac{a_3}{v_1} & 0 & 0 & 0 \\
-\frac{b_2}{ v_1}  & \frac{\alpha_2}{v_1} & 0 &0& 0 \\
\frac{a_3^2b_1b_2}{a_2 v_1^2} &-\frac{a_3^2b_1}{v_1^2} &  \frac{1}{a_2 } &-\frac{a_3b_3}{a_2 v_1} & \frac{a_3^2}{a_2 v_1}\\
 -\frac{a_3b_1b_2}{v_1^2} & \frac{a_2a_3b_1}{v_1^2} &  0 & \frac{b_3}{v_1} & -\frac{a_3}{v_1}\\
 \frac{a_2b_1b_2 }{v_1^2} &-\frac{a_2^2b_1}{v_1^2} &  0 &-\frac{b_2 }{v_1} & \frac{a_2 }{v_1}\end{pmatrix}
\end{equation}
Then the solution of equation \eqref{2c} is as follows:
\begin{equation}\label{11c}
\hat{n}_2 = \hat{P}_2\hat{\omega}_2,
\end{equation}
were
$$
\hat{n}_2 ^{T} = (n_{12}, n_{13}, n_{22}, n_{23}, n_{33}) ; \quad
$$
$$\hat{\omega}_2 = (-\dot{b}_{2}, (\dot{a}_{2}-b_1n_{11}), -\dot{b}_{1}, \frac{\dot{b}_{3}}{2}, -\frac{\dot{a}_{3}}{2}).
$$
Function \quad $n_{11}$, \quad  $a_a, \quad \beta_a$ \quad are arbitrary functions of \quad $u^0$, \quad that obey the condition \eqref{25f}.

\quad

3. \quad $a_3v_1 \ne 0 \Rightarrow  a_1=a_2=0 \Rightarrow$ the minor $\hat{B}^{-1}_{16}$ and its inverse matrix $\hat{P}_3 = \hat{B}^{-1}_{16}$ have the form:
\begin{equation}\label{9c}
\hat{B}_{16}=\begin{pmatrix} 0 & a_3 & 0 & 0 & 0
 \\ b_2 & b_3 & 0 & 0 & 0 \\ 0 & 0 & 0 & a_3 & 0 \\ b_1 & 0 & b_2 & b_3 & 0
\\0 & 0 & 0 &0 & a_3 \end{pmatrix},  \quad
\hat{P}_3=\begin{pmatrix}-\frac{b_3}{a_3b _2} & \frac{1}{b_3} & 0 & 0 & 0
 \\
\frac{1}{a_3} & 0 & 0 &0 & 0
\\
\frac{b_1b_3}{a_3b_2^2} &-\frac{b_1}{b_2^2} & -\frac{b_3}{b_2 a_3} &\frac{1}{b_2} & 0 \\
 0 &0 &\frac{1}{a_3} & 0 & 0
\\
0 & 0 &0 & 0 & \frac{1}{a_3}\end{pmatrix}
\end{equation}
Then the solution of equation \eqref{2c} is as follows:
\begin{equation}\label{8c}
\hat{n}_3 = \hat{P}_3\hat{\omega}_3,
\end{equation}
were
\begin{equation}\label{7c}
\hat{n}_3^{T} = ( n_{12}, n_{13}, n_{22}, n_{23}, n_{33}),\quad\hat{\omega}_3 = (-\dot{b}_{2}, (\dot{a}_{2}-b_1 n_{11}), -\dot{b}_{1}, 0, \frac{\dot{b}_3}{2}).
\end{equation}

\quad

4. \quad $\alpha_1v_2 \ne 0, \Rightarrow  v_1=0 \Rightarrow$ the minor $\hat{B}^{-1}_{24}$ and its inverse matrix $\hat{P}_4 = \hat{B}^{-1}_{24}$ have the form:
\begin{equation}\label{9c}
\hat{B}_{24}=\begin{pmatrix} a_1 & a_2 & a_3 & 0 & 0
 \\ 0 & a_1 & 0 & a_3 & 0 \\ 0 & b_1 & 0 & b_3 & 0  \\ 0 & 0 & a_1 & a_2 & a_3
\\0 & 0 & b_1 &b_2 & b_3 \end{pmatrix},  \quad
\hat{P}_{4}=\begin{pmatrix}
  \frac{1}{\alpha_1} & \frac{a_2b_3}{a_1 v_2} & -\frac{a_2a_3}{a_1 v_2} & \frac{a_3b_3}{a_1 v_2} &-\frac{a_3^2}{a_1 v_2} \\

0 & -\frac{b_3}{v_2} &\frac{a_3}{v_2} & 0 & 0 \\

 0 & 0 &  0 & -\frac{b_3}{v_2} & \frac{a_3}{v_2} \\

 0 & \frac{b_1}{v_2}&-\frac{a_1}{v_2} & 0 &  0\\

0 & \frac{b_1v_3}{v_2^2} & -\frac{a_1v_3}{v_2}  & \frac{b_1}{v_2} & - \frac{a_1}{v_2}\end{pmatrix}
\end{equation}
Then the solution of equation \eqref{2c} is as follows:
\begin{equation}\label{11c}
\hat{n}_{4} = \hat{P}_4\hat{\omega}_{4},
\end{equation}
were
\begin{equation}\label{10c}
\hat{n}_{4} ^{T} = (n_{11}, n_{12}, n_{13}, n_{23}, n_{33}), \quad
\hat{\omega}_{4} = (-\dot{b}_{2},-(\dot{b}_{1}+a_2 n_{22}),(\dot{a}_{1}-b_2n_{22}), \frac{\dot{b}_{3}}{2}, -\frac{\dot{a}_{3}}{2})
\end{equation}
Function \quad $n_{22},$ \quad $a_a, \quad \beta_a$ \quad are arbitrary functions of \quad $u^0$, \quad that obey the conditions \quad \eqref{25f} \quad and \quad $a_2\beta_3 = a_3\beta_2.$

\quad

5 . \quad $\alpha_2V_2 \ne 0, \Rightarrow  a_1 = V_1=0 \Rightarrow$ the minor $\hat{B}^{-1}_{44}$ and its inverse matrix $\hat{P}_5 = \hat{W}^{-1}_{44}$ have the form:

\begin{equation}\label{9c}
\hat{B}_{44}=\begin{pmatrix} 0 & a_2 & a_3 & 0 & 0
 \\ b_1 & b_2 & b_3 & 0 & 0 \\ 0 & 0 & 0 & a_3 & 0  \\ 0 & 0 & 0 & a_2 & a_3
\\0 & 0 & b_1 &b_2 & b_3 \end{pmatrix},  \quad
\hat{P_{5}}=\begin{pmatrix}
 -\frac{b_2}{b_1 a_2} &\frac{1}{b_1} & 0 & 0 & 0 \\
 \frac{1}{a_2} &  0 & 0 & \frac{b_3}{a_2b_1}& -\frac{a_3}{a_2b_1} \\
 0 & 0 & 0 & -\frac{b_3}{a_3b_1} & \frac{1}{b_1}\\
 0 & 0 & \frac{1}{a_3}& 0 & 0\\
0 & 0 & -\frac{a_2}{a_3^2} & \frac{1}{a_3} & 0\end{pmatrix}
\end{equation}

Then the solution of equation \eqref{2c} is as follows:
\begin{equation}\label{11c}
\hat{n}_{5} = \hat{P}_2\hat{\omega}_{5},
\end{equation}
were
\begin{equation}\label{11f}
\hat{n}_{5} ^{T} = (n_{11},n_{12}, n_{13}, n_{23}, n_{33}) ; \quad
\hat{\omega}_{5} = (-\dot{b}_2,\dot{\alpha}_2,-(\dot{b}_{1}+a_2 n_{22}),\frac{\dot{b}_{3}}{2}, -\frac{\dot{\alpha}_{3}}{2})
\end{equation}
Function \quad $n_{22},$ \quad $a_a, \quad b_a$ \quad are arbitrary functions of \quad $u^0$, \quad that obey the conditions \quad \eqref{25f} \quad and \quad $a_2b_3 = a_3b_2.$

\quad

6) \quad $a_3v_2 \ne 0, \quad v_1=0, \Rightarrow  a_1=a_2=b_2 = 0$.
From the condition \quad \eqref{25f} it follows:
$$
a_3 = c\cos{2\varphi}, \quad   b_3 = c\sin{2\varphi},
$$
$\varphi$ is an arbitrary function of $u^0$.
The minor \quad  $\hat{B}^{-1}_{46}$ \quad  and its inverse matrix \quad  $\hat{P}_{6} = \hat{B}^{-1}_{46}$ \quad  have the form:
\begin{equation}\label{9c}
\hat{B}_{64}=\begin{pmatrix} 0 & 0 & c\cos{\varphi} & 0 & 0\\
b_1 & 0 &  & 0 & 0 \\
 0 & 0 & 0 & c\cos{\varphi} & 0 \\
  0 &b_1 & 0 & c\sin{\varphi} & 0
\\0 & 0 & 0 & 0 & c\cos{\varphi} \end{pmatrix},
 \hat{\Omega_{6}}=\begin{pmatrix}
 -\frac{\sin{\varphi}}{b_1 \cos{\varphi}} &\frac{1}{b_1} & 0 & 0 & 0 \\
 0 &  0 & -\frac{\sin{\varphi}}{b_1 \cos{\varphi}} & \frac{1}{b_1}& 0 \\
 \frac{1}{c\cos{\varphi}}& 0 & 0 & 0 & 0\\
 0 & 0 & \frac{1}{c\cos{\varphi}}& 0 & 0\\
0 & 0 & 0 & 0 & \frac{1}{c\cos{\varphi}}\end{pmatrix}.
\end{equation}
Then the solution of equation \eqref{2c} is as follows:
\begin{equation}\label{11c}
\hat{n}_{6} = \hat{P}_{6}\hat{\omega}_{6},
\end{equation}
were
$$
\hat{n}_{6} ^{T} = (n_{11},n_{12}, n_{13}, n_{23}, n_{33}) ; \quad
\hat{\omega}_{6} = (0, 0, -\dot{b}_{1},0, c\dot{\varphi}\cos{\varphi}).
$$
Function \quad $n_{22}$ \quad \quad$ b_1, \quad\varphi$ \quad are arbitrary functions of \quad $u^0$.

\quad

7. \quad $a_1v_3 \ne 0$ $\Rightarrow v_1=v_2=0,$ \quad otherwise, we get a solution equivalent to the previous ones. As \quad $v_3\ne 0 \Rightarrow$ \quad $a_3=b_3=0.$\quad The minor \quad $\hat{B}_{26}$ \quad and its inverse matrix \quad $\hat{P}_{7} = \hat{B}^{-1}_{26}$ \quad have the form:
\begin{equation}\label{9c}
\hat{B}_{26}=\begin{pmatrix} \alpha_1 & \alpha_2 & 0 & 0 & 0
 \\ 0 & \alpha_1 & 0 & a_2 & 0 \\ 0 & b_1 & 0 & b_2 & 0 \\ 0 & 0 & \alpha_1 & 0 & \alpha_2
\\0 & 0 & b_1 & 0 & b_2 \end{pmatrix},  \quad \hat{P}_{7}=
\begin{pmatrix} \frac{1}{\alpha_1} & -\frac{\alpha_2b_2}{\alpha_1v_3} & \frac{\alpha_2^2}{\alpha_1v_3} & 0 & 0
 \\ 0 & \frac{b_2}{v_3} & -\frac{\alpha_2}{v_3} & 0 & 0 \\ 0 & 0 & 0 & \frac{b_2}{v_3} & -\frac{\alpha_2}{v_3} \\ 0 & -\frac{b_1}{v_3} & \frac{\alpha_1}{v_3} & 0 & 0
\\0 & 0 & 0 & -\frac{b_1}{v_3} & \frac{\alpha_1}{v_3} \end{pmatrix}.
\end{equation}
Then the solution of equation \eqref{2c} is as follows:
\begin{equation}\label{8c}
\hat{n}_{3a} = \hat{P}_{7}\hat{\omega}_{7}.
\end{equation}
were
$$
\hat{n}_{7}^{T} = (n_{11}, n_{12}, n_{13}, n_{22}, n_{23});$$ \quad
$$
\hat{\omega}_{7}^{T} = (-\dot{b}_{2}, -\dot{b}_{1}, \dot{a}_{1}, 0, 0).
$$

\quad

8. \quad $a_2v_3 \ne 0$. $\Rightarrow a_1=v_1=v_2=0,$ \quad otherwise we get a solution equivalent to the previous ones. As \quad $v_3\ne 0 \Rightarrow$ \quad $\alpha_3=b_3=0.$\quad The minor \quad $\hat{B}_{64}$\quad and its inverse matrix \quad $\hat{P}_8 = \hat{B}^{-1}_{64}$ \quad have the form:
\begin{equation}\label{9c}
\hat{B}_{64}=\begin{pmatrix} 0 & \alpha_2 & 0 & 0 & 0
\\ b_1 & \alpha_2 & 0 & 0 & 0
\\ 0 & 0 & 0 & a_2 & 0
\\ 0 & 0 & 0 & 0 & \alpha_2
\\0 & 0 & b_1 & 0 & b_2 \end{pmatrix},  \quad \hat{P}_{8}=
\begin{pmatrix} -\frac{b_2}{a_2 b_1} & -\frac{1}{b_1} & 0 & 0 & 0 \\ \frac{1}{a_2} & 0 & 0 & 0 & 0 \\

  0 & 0 & 0 & -\frac{b_2}{b_1a_2} & \frac{1}{b_1} \\

   0 & 0 & \frac{1}{a_2} & 0 & 0 \\

0 & 0 & 0 & \frac{1}{a_2} & 0 \end{pmatrix}.
\end{equation}
Then the solution of equation \eqref{2c} is as follows:
\begin{equation}\label{8c}
\hat{n}_8 = \hat{P}_8\hat{\omega}_8.
\end{equation}
were
$$
\hat{n}_{8}^{T} = (n_{11}, n_{12}, n_{13}, n_{22}, n_{23}), \quad
\hat{\omega}_{8}^{T} = (-\dot{b}_{2}, -\dot{b}_{1}, 0, 0, 0).
$$
Function \quad $n_{33},$  $a_2 \quad \beta_1, \quad\beta_2 $ \quad are arbitrary functions of \quad $u^0$, \quad that obey the condition \eqref{25f}.

\quad

II. \quad $rank||\hat{B}||<5$

9.\quad $v_a =0 $.\quad Let us represent the system of Maxwell's equations in the form:
\begin{equation}\label{9c}\hat{Q}\hat{n}_I = \hat{\omega}_I,\end{equation}
were
$$\hat{Q} = \begin{pmatrix} a_1 & a_2 & a_3 & 0 & 0 & 0 \\ 0 &  a_1 & 0 & a_2 & a_3 &0
\\0 & 0 & a_1 & 0 & a_2 & a_3
 \\ b_1 & b_2 & b_3 & 0 & 0 & 0 \\0 & b_1 & 0 & b_2 & b_3 & 0
\\0 & 0 & b_1 & 0 & b_2 & b_3\end{pmatrix},$$

$$
\hat{\omega}_I = (\hat{\omega}_\beta, \hat{\omega}_\alpha); \quad \hat{\omega}_\beta^T =(-\dot{b}_{2}, -\dot{b}_{1}, \frac{\dot{b}_{3}}{2}), \quad \hat{\omega}_\alpha^T = (\dot{a}_{2}, \dot{a}_{1}, -\frac{\dot{a}_{3}}{2})
$$
$$
\hat{n}_I = (\hat{n}_\alpha, \hat{b}_\alpha); \quad \hat{n}_\alpha^T =(n_{11},n_{12},n_{13}), \quad \hat{n}_\beta^T =(n_{22},n_{23},n_{33})
$$
Consider all possible options.

\quad

a) $ a_1\ne 0 \Rightarrow b_a=\frac{\alpha_a b_1}{\alpha_1}$.
Maxwell's equations \eqref{9c} take the form:
$$
 \hat{B}_I \hat{n}_\alpha = (\hat{\omega}_\beta -\hat{B}_{II} \hat{n}_\beta) \Rightarrow
 \hat{n}_\alpha = \hat{B}_I^{-1}(\hat{\omega}_\beta -\hat{B}_{II} \hat{n}_\beta) ,$$
 $$
b_1 \hat{B}_I \hat{n}_\alpha = a_1\hat{\omega}_\alpha -b_1\hat{B}_{II} \hat{n}_\beta  \Rightarrow b_1\hat{\omega}_\beta - a_1\hat{\omega}_\alpha  =0  \Rightarrow
 $$
 \begin{equation}\label{18c}
   \left\{\begin{array}{rl}
a_1\dot{a}_2 +b_1\dot{b}_2 =0, \\
a_1\dot{a}_3 +b_1\dot{b}_3 =0, \\
a_1\dot{a}_1 +b_1\dot{b}_1 = 0. \\
\end{array}\right.\end{equation}
Here:
$$
 \hat{B}_I = \begin{pmatrix} a_1 & a_2 & a_3 &  \\ 0 &  a_1 & 0
\\0 & 0 & a_1
 \end{pmatrix}, \hat{B}_I^{-1}= \begin{pmatrix} \frac{1}{a_1} & -\frac{a_2}{a_1^2} & -\frac{a_3}{a_1^2} &  \\ 0 &  \frac{1}{a_1} & 0
\\0 & 0 & \frac{1}{a_1}
 \end{pmatrix}, \hat{B}_{II} = \begin{pmatrix} 0 & 0 & 0 &  \\ a_2 &  a_3 & 0
\\0 & a_2 & a_3,
 \end{pmatrix}
 $$
From the last equation of the system \eqref{18c} it follows:
$$ a_1 = e_0 \sin\varphi, \quad b_1 = e_0 \cos\varphi, \quad e_0 = const. $$
Thus \quad $ b_2 = a_2\frac{\cos \varphi}{\sin \varphi},\quad b_3 = a_3\frac{\cos \varphi}{\sin \varphi}$,\quad
and from the previous equations, it follows:
$$
a_a = e_0q_a\sin\varphi, \quad b_a = e_0q_a \cos\varphi, \quad q_a = const, \quad q_1 =1. $$
Then matrices \quad $\hat{B}_I,\quad \hat{B}_I^{-1},\quad \hat{B}_{II}$ \quad and lines \quad $\hat{\omega}^T$ take the form:
$$
 \hat{B}_I = \hat{w}_1\sin{\varphi},\quad \hat{B}_I^{-1}= \frac{1}{\sin{\varphi}}\hat{w}^{-1}_1 ,\quad \hat{B}_{II} = \hat{w}_2 sin{\varphi}.
 $$
 $$
 \hat{w}_1=\begin{pmatrix} 1 & q_2 & q_3 &  \\ 0 &  1 & 0
\\0 & 0 & 1
 \end{pmatrix}, \quad
 \hat{w}^{-1}_1 = \begin{pmatrix} 1 & -q_2 & -q_3 \\ 0 &  1 & 0
\\0 & 0 & 1 \end{pmatrix},\quad
\hat{w}_2 = \begin{pmatrix} 0 & 0 & 0 &  \\ q_2 &  q_3 & 0
\\0 & q_2 & q_3,
 \end{pmatrix}
 $$
 $$
\hat{\omega}^T_\beta = \dot{\varphi} \hat{c}^T = \dot{\varphi}\sin{\varphi}(q_2, 1, -\frac{q_3}{2}), $$
Then the solution of equation \eqref{2c} is as follows:
$$
\hat{n}_\alpha = \hat{w}^{-1}(\dot{\varphi}\hat{c} -\hat{q} \hat{n}_\beta)
$$

\quad

b)\quad $ a_1=0 \quad \Rightarrow \quad a_2 \ne 0$.\quad Let us use the previous results, in which the indices 1 and 2 are reversed: $1 \Leftrightarrow2$. The solution of Maxwell's equations has the form:
$$
\hat{n}_\alpha = \hat{w}^{-1}(\dot{\varphi}\hat{c}-\hat{q} \hat{n}_\beta)
$$
$$
 \hat{n}_\alpha^T =(n_{22},n_{12},n_{23}), \quad \hat{n}_\beta^T =(n_{11},n_{13},n_{33}),
 $$
 $$
 \hat{w}^{-1} = \begin{pmatrix} 1 & 0 & -q \\ 0 &  1 & 0
\\0 & 0 & 1 \end{pmatrix},\quad
\hat{q} = \begin{pmatrix} 0 & 0 & 0 &  \\ 0 &  q & 0
\\0 & 0 & q,\end{pmatrix},  \quad  \hat{c}^T = (0, 1, -\frac{q}{2}).
 $$
 $$ a_2 = e_0\sin\varphi, \quad b_2 = e_0 \cos\varphi,\quad a_3 = e_0q \sin\varphi, \quad b_3 = e_0q \cos\varphi,\quad  q = const, \quad \varphi=\varphi(u^0). $$

 \quad

 c)$\quad a_3 \ne 0.$ \quad The solutions, which are not equivalent to the previous ones, can be obtained under the conditions\quad $a_1 =a_2 = 0 \Rightarrow b_1=b_2 =0$.\quad From  Maxwell's equations it follows:
$$
a_3 n_{13}=a_3 n_{23}=0,\quad a_3 n_{33}=\frac{\dot{b}_3}{2}, \quad b_3 n_{33}=-\frac{\dot{a}_3}{2}\Rightarrow
a_3 \dot{a}_3 + b_3\dot{b}_3=0..
$$
The solution has the form:
$$
 n_{33}=\dot{\varphi}, \quad n_{13}= n_{23}=a_1 = a_2  =b_1 = b_2 = 0,\quad a_3 =q\cos{2\varphi}, \quad b_3 = q\sin{2\varphi}.
$$
Functions \quad $ \varphi,\quad n_{11}, \quad  n_{12}, \quad  n_{22}$ - are arbitrary functions on $u^0$.

\quad

\section{Conclusion}
In the previous works [\cite{34} \cite{35},\cite{ 36} ] ], all non-equivalent solutions of Maxwell's empty equations for admissible electromagnetic fields in homogeneous space-time metrics of all types according to Bianchi's classification, except type $VIII $, were found. The present work completes the first stage of the classification problem formulated in the introduction. The next step is the classification of the corresponding exact solutions of the Einstein-Maxwell equations. All solutions obtained in the completed classification have a form suitable for further use and have sufficient arbitrariness so that the Einstein-Maxwell equations have nontrivial solutions. The use of the triad of frame vectors (see \cite{32}) allows us to reduce the Einstein-Maxwell equations with the energy-momentum tensor of the admissible electromagnetic field to an overcrowded system of ordinary differential equations. To perform the classification, we need to study the coexistence conditions of these systems of equations. It is possible to use additional symmetries of homogeneous spaces and admissible electromagnetic fields (see \cite{31}). In the future, we will start to solve this classification problem.

\quad

FUNDING: The work is supported by Russian Science Foundation, project number N 23-21-00275.

INSTITUTIONAL REVIEW BOARD STATEMENT: Not applicable.

INFORMED CONSENT STATEMENT: Not applicable.

DATA AVAILABILITY STATEMENT: The data that support the findings of this study are available within the article.

CONFLICTS OF INTEREST: The author declares no conflict of interest.


\begin{thebibliography}{99}

\bibitem{1} Stackel. P. Uber die intagration der Hamiltonschen differentialechung mittels separation der variablen.  {\em Math. Ann.}
{\bf 1897}. {\em 49}, (145-147 pp.);

\bibitem{2}
 Eisenhart L.P. Separable systems of stackel. {\em Ann.Math.} {\bf 1934}, {\em 35}, (284-305 pp).;

\bibitem{3}
Levi-Civita T. Sulla Integraziome Della Equazione Di Hamilton-Jacobi Per Separazione Di Variabili. {\em Math.Ann.} {\bf 1904} {\em 59}, (383-397 pp.);

\bibitem{4}
Jarov-Jrovoy M.S. Integration of Hamilton-Jacobi equation by complete separation of variables method. {\em J.Appl.Math.Mech.}. {\bf 1963}, {\em27}, {\em No 6},  (173-219 pp).https://doi.org/10.1016/0021-8928(63)90122-9;

\bibitem{5}
Carter B. New family of Einstein spaces. {\em Phys.Lett.}
{\bf 1968}, {\em A.25}, {\em No 9} (399-400pp.), doi.org/10.1016/0375-9601(68)90240-5

\bibitem{6}
 Shapovalov V.N., Symmetry and separation of variables in the Hamilton-Jacobi equation. {\em Sov. Phys.J.}.  {\bf 1978}, {\em 21}, 1124-1132pp.      doi: 10.1007/BF00894560;

\bibitem{7}
 Shapovalov V.N., Stackel`s spaces. {\em Sib. Math. J.} {\bf 1979}, {\em 20}, (1117-1130pp.), doi: org/10.1007/BF00971844;

\bibitem{8} W. Miller. Symmetry And Separation Of Variables. {\em Cambridge University Press:Cambridge}. {\bf1984}, (318 p.p.);

\bibitem{9}
Obukhov V.V. Hamilton-Jacobi equation for a charged test particle in the Stackel space of type (2.0). {\em Symmetry}, {\bf12}, {\bf 2020}, 12891291. doi: 10.3390/sym12081289.

\bibitem{10}
Obukhov V.V.  Hamilton-Jacobi equation for a charged test particle in the  Stackel space of type (2.1).{\em Int. J. Geom. Meth. Mod. Phys}, {\bf 2020}, {\bf 17}, {\em 14}, {2050186}. doi: 10.1142/S0219887820501868.

\bibitem{11} Obukhov V.V. Separation of variables in Hamilton-Jacobi and       Klein-Gordon-Fock equations for a charged test particle in the stackel spaces of type (1.1). {\em Int. J. Geom. Meth. Mod. Phys}.{\bf 2021}, {\bf18}, {\em 03}, (2150036); doi:10.1142/S0219887821500365

\bibitem{12} Mitsopoulos A., Tsamparlis M., Leon G. A.Paliathanasis. A.Paliathanasis. New conservation laws and exact cosmological solutions in Brans-Dicke cosmology with an extra scalar field. {\em Symmetry}. {\bf 2021}, 13(8):1364, doi: 10.3390/sym13081364

\bibitem{13}
Claudio Dappiaggi, Benito A. Juarez-Aubry, Alessio Marta Ground. State for the Klein-Gordon field in anti-de Sitter spacetime with dynamical Wentzell boundary conditions. {\it Phys. Rev. D.} {\bf 2022}, {\em 105}, 105017 doi.org/10.1103/PhysRevD.105.105017.

\bibitem{14}
Francisco Astorga, J Felix Salazar and Thomas Zannias
On the integrability of the geodesic flow on a Friedmann-Robertson-Walker spacetime  {\it Physica Scripta}, {\bf 2018}, {\bf 93}, {\em 8}, 93 085205
doi. 10.1088/1402-4896/aacd44

\bibitem{15}
Capozziello S., De Laurentis M., Odintsov D. Hamiltonian dynamics and Noether symmetries in extended gravity cosmology. {\em Eur.Phys.J.} {\bf 2012}, {\em C72}, 2068 (22 pp.), doi: 10.1140/epjc/s10052-012-2068-0;

\bibitem{29a}
Odintsov, S.D. Editorial for Feature Papers 2021-2022. {\em Symmetry}. {\bf2023}, {\em 15}, 32. doi.org/10.3390/sym15010032

\bibitem{15a} Kibaroglu Salih, Cebecioglu Oktay.   Generalized cosmological constant from gauging Maxwell-conformal algebra, {\em Phys.Lett.B}, {\bf2020}, {\em 803}, 135295. doi.org/10.1016/j.physletb.2020.135295

\bibitem{15b}
Cebecioglu Oktay., Kibaroglu Salih. Maxwell-modified metric affine gravity,
{\em The European Physical Journal}. {\bf2021}, {\em 81}, {\em 10}, (900). doi.org/10.1140/epjc/s10052-021-09685-6

\bibitem{15c}
Medine Ildes and Metin Arik. Analytic solutions of Brans-Dicke cosmology: Early inflation and late time accelerated expansion,{\em International Journal of Modern Physics}. {\bf2023}, {\em32}, {\em 1}, (2250131).doi.org/10.1142/S0218271822501310

\bibitem{16}
Nojiri Shin'ichi, Odintsov Sergei D., Faraoni Valerio, Searching for dynamical black holes in various theories of gravity. {\em Phys.Rev.D}. {\bf 2021}, {\em103}, 4, 044055. doi:10.1103/PhysRevD.103.044055

\bibitem{17}
 Epp V. Pervukhina O. The Stormer problem for an aligned rotator. {\em MNRAS},{ \bf 2018} {\em 474}, (5330-5339)  doi.org/10.1093/mnras/stx3102.

\bibitem{17a}
Epp V., Masterova M.A. Effective potential energy for relativistic particles in the field of inclined rotating magnetized sphere. {\em Astrophys Space Sci}, {\bf 2014}, {\em 353}, (473-483). doi.org/10.1007/s10509-014-2066-9.

\bibitem{18} Kumaran Y.; Ovgun A. Deflection angle and shadow of the
reissner-nordstrom black hole with higher-order magnetic correction in
einstein-nonlinear-maxwell fields. {\em Symmetry.}  {\bf 2022}, {\em14}, 2054. https://doi.org/10.3390/sym14102054

\bibitem{19}
Osetrin K.; Osetrin E. Shapovalov wave-like spacetimes.
{\em Symmetry} {\bf 2020}, {\em 12}, 1372.
https://doi.org/10.3390/SYM12081372.

\bibitem{23}
Shapovalov, A.V., Shirokov I.V. Noncommutative integration method for linear partial differential equations. functional algebras and dimensional reduction. {\em  Theoret. And Math. Phys.} {\bf 1996}, {\em 106:1}, (1-10). doi.org/10.4213/tmf1093.

\bibitem{23a}
Petrov A. Z. Einstein Spaces, {Pergamon Press,\em Oxford}, {\bf 1969}.

\bibitem{24}
Breev A.; Shapovalov A.; Gitman D. Noncommutative eduction of Nonlinear Schredinger Equation on Lie Groups. {\em Universe.}
{\bf 2022}, 8, 445. doi.org/10.3390/universe8090445

\bibitem{25}
 Breev, A.I., Shapovalov, A.V. Non-commutative integration of the Dirac equation in homogeneous spaces. {\em Symmetry } {\bf 2020}, {\em 12}, 1867.

\bibitem{26}
 Breev, A.I.; Shapovalov, A.V. Yang--Mills gauge fields conserving the symmetry algebra of the Dirac equation in a homogeneous space. {\em J. Phys.: Conf. Ser.} {\bf 2014}, 563, 012004.

\bibitem{27}
 Magazev A.A., Boldyreva M.N. Schrodinger equations in electromagnetic fields:
symmetries and noncommutative integration, {\em Symmetry}. {\bf 2021}, {\em 13}, (1527).
https://doi.org/10.3390/sym13081527

\bibitem{20o}
Osetrin E.; Osetrin K.; Filippov A.
 Plane Gravitational Waves in Spatially-Homogeneous Models of type-(3.1) {Stackel} Spaces.
 {\em Russian Physics Journal} {\bf 2019}, {\em 62}, ~292-301.
https://doi.org/10.1007/s11182-019-01711-1.

\bibitem{21o}
Osetrin K., Osetrin E. and Osetrina E. Geodesic deviation and tidal acceleration in the gravitational wave of the Bianchi type IV universe. {\em Eur. Phys. J. Plus}. {\bf 2022}, {\em 137}, (856). https://doi.org/10.1140/epjp/s13360-022-03061-3

\bibitem{22o}
Osetrin, K.; Osetrin, E.; Osetrina, E.
Gravitational wave of the {Bianchi VII} universe: Particle
trajectories, geodesic deviation and tidal accelerations.
{\em Eur. Phys. J. C} {\bf 2022}, {\em 82},~1--16.
doi.org/10.1140/epjc/s10052-022-10852-6.

\bibitem{22A}
Md Nur Alam, Cemil Tunc Constructions of the optical solitons and other solitons to the conformable fractional Zakharov-Kuznetsov equation with power law nonlinearity. {\em Journal of Taibah University for Science}.{\bf 2020}. {\em 14:1}, (94-100), DOI: 10.1080/16583655.2019.1708542

\bibitem{22B}
Md Al-Asad, Md Nur Alam, Cemil Tunc, MMA Sarker. Heat transport exploration of free convection flow inside enclosure having vertical wavy walls. {\em J. Appl. Comput. Mech.} {\bf 2021}, {\em 7(2)}, (520-527).


\bibitem{28}
Magazev A.A. Integrating Klein-Gordon-Fock equations in an extremal
electromagnetic field on Lie groups. {\em Theor.and Math.Phys.}, {\bf 2012}, {\em 173:3}, (1654-1667). doi: 10.1007/s11232-012-0139-x, arxiv.org/abs/1406.5698;

\bibitem{29}
Obukhov V.V. Algebra of symmetry operators for Klein-Gordon-Fock Equation. {\em Symmetry}. {\bf2021}, {\em 13},  727 (15p.). https://doi.org/10.3390/sym13040727.

\bibitem{30}
Obukhov V.V. Algebra of the symmetry operators of the Klein-Gordon-Fock equation for the case when groups of motions $G_3$ act
transitively on null subsurfaces of spacetime. {\em Symmetry}. {\bf2022}, {\em 14}, (346). https://doi.org/10.3390/sym14020346

\bibitem{31}
Obukhov V.V. Algebras of integrals of motion for the Hamilton-Jacobi and Klein-Gordon-Fock equations in spacetime with a four-parameter groups of motions in the presence of an external electromagnetic field {\em J. Math. Phys.} {\bf 2022}, {\em 63}, Issue 2.
https://doi.org/10.1063/5.0080703

\bibitem{34}
Obukhov V.V. Maxwell Equations in Homogeneous Spaces for Admissible Electromagnetic Fields.{\em Universe.} {\bf 2022}, {\em 8}, (245). https://doi.org/10.3390/universe8040245

\bibitem{35}
Obukhov, V.V. Maxwell Equations in Homogeneous Spaces with Solvable Groups of Motions. {\em Symmetry.} {\bf 2022}, {\em 14}, (2595).https://doi.org/10.3390/sym14122595

\bibitem{36}
 Obukhov V. V. Exact Solutions of Maxwell Equations in Homogeneous Spaces with the Group of Motions G3(IX). {\em Axioms.}
{\bf 2023}, {\em 12}, {\em 135}. doi.org/10.3390/axioms12020135

\bibitem{32}
 Landau L.D., Lifshits E.M. Theoretical physics, Field theory. 7th ed. Moskow.(vol. II): Science. Ch. ed. Phys.-Math. lit., {\bf 1988}. - (512 p). ISBN 5-02-014420-7

\end{thebibliography}
\end{document}